\documentclass[aps,prb,twocolumn,floatfix,floats,showpacs,superscriptaddress]{revtex4}
\usepackage{graphicx,latexsym}
\usepackage{dcolumn}
\usepackage{amsmath}

\begin{document}
\title {Effects of a transversely polarized electric field\\
on the quantum transport in narrow channels}
\author{C. S. Chu}
\affiliation{Department of Electrophysics, National Chiao Tung
University, Hsinchu 30010, Taiwan}
\author{C. S. Tang}
\affiliation{Physics Division, National Center for Theoretical
        Sciences, P.O.\ Box 2-131, Hsinchu 30013, Taiwan}
\begin{abstract}
The quantum transport in a narrow channel (NC) is studied in the
presence of a time-dependent delta-profile electric field.
 The electric field is
taken to be transversely polarized, with frequency $\omega$, causing
inter-subband and inter-sideband transitions. Suppression in the dc
conductance $G$ is found, which escalates with the chemical
potential. There are structures in $G$ which are related to the
quasi-bound states (QBS) features. Major dip, and dip-and-peak,
structures occur when an incident electron makes transition to a
subband edge by absorbing or emitting one, and two, $\hbar\omega$,
respectively. Structures associated with three $\hbar\omega$
processes are recognized. The QBS are closely associated with the
singular density of states (DOS) at subband bottoms.  Our results
indicate that, due to this singular features of the DOS, the
interaction of the electron with the electric field has to be
treated beyond finite order perturbation.
\end{abstract}
\maketitle

Quantum transport in NC has received a lot of attention in recent
years. These channels can be realized experimentally within a
split-gate configuration.\cite{wee88,wha88}  The channels connect
adiabatically on each side to a two-dimensional electron gas. Energy
levels of the channels are quantized into one-dimensional subbands
which give rise, in the ballistic regime,  to a quantized
$G$.\cite{wee88,wha88}  The singular feature of the DOS near a
subband edge leads to impurity-induced dip structures in $G$ when
the scatterer is
attractive.\cite{chu89,bag90,tek91,nix91,lev92,tak92,kun92}  These
dip structures are associated with the formation of impurity-induced
QBS.\cite{bag90}

More recently, there are growing interest in the time-dependent
phenomena in quantum point contact (QPC) systems.
\cite{gri94,qin93,fen93,wys93,gor94,jan94} Firstly, the optical
absorption coefficient of a QPC has been calculated\cite{gri94}, up
to second-order in the electron-photon coupling, and a proposal is
made that the optical absorption in the QPC can be used to
characterize the lateral confining potential of the QPC.  Secondly,
the photon-assisted quantum transport in QPC systems
\cite{qin93,fen93,wys93,gor94,jan94} has been studied while
ministeps are predicted to appear in $G$ versus $\mu$.\cite{fen93}

On the other hand, QBS features are found in $G$ when a point
barrier oscillates in a narrow channel.\cite{bag92}  These features,
that the dc conductance $G$ exhibits dip, or peak structures when
the chemical potential $\mu$ is at $n\hbar\omega$ above the
threshold energy of a subband, persists even in the case of a
finite-range oscillating barrier, including the case when the
barrier range $a \gg \lambda_{F}$.\cite{tan95}  The oscillating
barrier is uniform in the transverse direction and does not induce
inter-subband transitions.  However, the presence of the oscillating
barrier, as long as it has a longitudinal profile, breaks the
longitudinal translational invariance.  Thus the electrons are
relieved from conserving the longitudinal momentum, making the
inter-sideband transitions possible.  The sideband index $n$ labels
those electrons which net energy change is $n\hbar\omega$, as a
result of interacting with the oscillating barrier.  The physical
origin of these QBS, as pointed out by Bagwell and Lake\cite{bag92}
to be essentially similar to that in the impurity-induced QBS, is
associated with the singular DOS at subband bottoms.\cite{fnt1} From
the above understanding, the QBS features are induced by an
oscillating barrier with a longitudinal profile.  Hence it is
interesting to ask whether a transversely polarized time-dependent
electric field, with a longitudinal profile, could induce similar,
or richer, QBS features?

In this paper, we study the quantum transport in a NC which is acted
upon by an electric field.  The transverse confinement of the NC is
modelled by a quadratic potential\cite{but90} and the transversely
polarized time-dependent electric field
$\displaystyle{\vec{E}(\vec{x},t)=E(x)cos(\omega t) \, \hat{y}}$ is
taken to have a delta-profile.  The effect of this electric field is
represented by a potential $\displaystyle{eE_{0}y \, \delta(x) \,
cos\omega t}$, where $-e$ is the charge of an electron and $E_{0}$
is essentially the integral of $E(x)$ with respect to the
longitudinal coordinate $x$.  Even though our delta-type profile
electric field does not correspond to the experimental situation,
 we still expect
that features obtained in this calculation  persist in the more
realistic case of a finite-range-profile electric field.  The
reasons being, firstly, that this model has incorporated the
essential processes: the inter-subband, the intra-subband, and the
inter-sideband transitions.  Secondly, our approach has a
renormalization feature which is important in narrow channels,
especially when QBS should exist.\cite{kun92}  Thirdly, as discussed
above, our recent study\cite{tan95} shows that the features of the
QBS found in the point-oscillating-barrier\cite{bag92} persist even
when the oscillating barrier has a large-finite-range profile.  This
demonstrates that the key, and the only significant, role of the
delta profile is to break the conservation of the longitudinal
momentum, allowing the electron to undertake an energy change of
$n\hbar\omega$, other than the subband energy difference.   Even
though the potential is not the same as the delta-profile electric
field in this paper, the results\cite{tan95} hint that the
delta-profile assumption is a reasonable simplification to the more
realistic case of a finite-profile electric field.  The results thus
obtained should be qualitatively sound.  We intend to extend, in the
near future, our calculation to the case of a finite-profile
electric field.

Choosing the energy unit $E^{*} = \hbar^{2} k_{F}^{2} / 2m^{*}$, the
length unit $a^{*} = 1 / \! k_{F}$, the time unit $t^{*} = \hbar /
E^{*}$, and $E_{0}$ in units of $E^{*} / e$,
the dimensionless two-dimensional Schr\"{o}dinger
equation becomes
\begin{widetext}
\begin{equation}
 \left[-\nabla^{2}+\omega_{y}^{2} \, y^{2}+E_{0} \, y \,
\delta(x) \, cos\omega t
\, \right] \Psi(\vec{x},t) = i\frac{\partial}{\partial t} \,
\Psi(\vec{x},t).
\end{equation}
Here $k_{F}$ is a typical Fermi wavevector of the reservoir.
The transverse energy levels are quantized, given by  $\varepsilon_{n}=
(2n+1) \,
 \omega_{y}$.
For a nth subband electron incident along $\hat{x}$, and with energy
$\mu$, the scattering wave function is in the form
\begin{equation}
\Psi_{n}^{(+)}(\vec{x},t) = \left\{ \begin{array}{ll}
\phi_{n}(y)e^{ik_{n}(\mu )x}e^{-i\mu t} +
\displaystyle \sum_{n',m'}r_{nn'}(m')\phi_{n'}(y)e^{-ik_{n'}(\mu
+m'\omega )x}e^{-i(\mu +m'\omega)t} &\quad\mbox{$x<0$,} \\
\displaystyle \sum_{n',m'}t_{nn'}(m')\phi_{n'}(y)e^{ik_{n'}(\mu
+m'\omega )x}e^{-i(\mu +m'\omega)t} &\quad\mbox{$x>0$,}
    \end{array}
    \right.
\end{equation}
\end{widetext}
where $n'$ and $m'$ are the final subband and sideband indices,
respectively.  A positive (negative) integer $m'$ corresponds to the
case
when the reflected or the transmitted wave has absorbed (emitted) a
net energy of $m'\hbar\omega$.  The wavevector $k_{n}(\mu) =
\sqrt{\mu
-(2n+1)\omega_{y}}$ is the effective wavevector for the electron with
energy $\mu$ and in the nth subbands.

Matching the wave functions at $x=0$ and given the expression of the
matrix element
\begin{equation}
<l\mid y\mid n'> = {1\over \sqrt{2\omega_{y}}}\left[\sqrt{n'}\delta_{l,n'-1} +
\sqrt{n'+1} \, \delta_{l,n'+1}\right],
\end{equation}
we obtain the equations for the reflection coefficients $r_{nl}(m)$
and the transmission coefficients $t_{nl}(m)$
\begin{equation}
 t_{nl}(m) - r_{nl}(m) = \delta_{m,0} \, \delta_{n,l} \, ,
\end{equation}
and
\begin{eqnarray}
k_{l}(\mu+m\omega) t_{nl}(m) &=& \frac{E_{0}}{4i}
\sum_{n',m'} \left[ \delta_{m', m+1}+ \delta_{m', m-1}\right] \nonumber  \\
&&  \times <l\mid y\mid n'> t_{nn'}(m') \nonumber
\\ && + \delta_{m,0} \ \delta_{n,l} \, k_{n}(\mu).
\end{eqnarray}

 To show the contribution to $G$
from various electron-photon processes, we write $G = (2e^2/h) \,
\sum_{n=0}^{N} \, G_{n}$ , where N+1 is the number of propagating
incident subbands. The contribution to $G$ from the nth subband
incident electrons is $G_{n}=\sum_{n'=0}^{N}\sum_{m'} G_{nn'}^{m'}$
\ ,where \ $\displaystyle{G_{nn'}^{m'}}$ is related to the
transmission coefficients, given by
\begin{equation}
G_{nn'}^{m'} = \frac{k_{n'}(\mu+m'\omega)}{k_{n}(\mu)} \mid
t_{nn'}(m')\mid^{2}.
\end{equation}
Here, $k_{n}(\mu)=\sqrt{\mu-(2n+1)\omega_{y}}$ is the effective
wavevector for the electron with energy $\mu$ in the nth subband.
The conservation of the dc current, given by the condition
$G_{n}=Re\left[ t_{nn}(0)\right]$, is used to check our results.
\begin{figure}[t]
      \includegraphics[width=0.40\textwidth,angle=0]{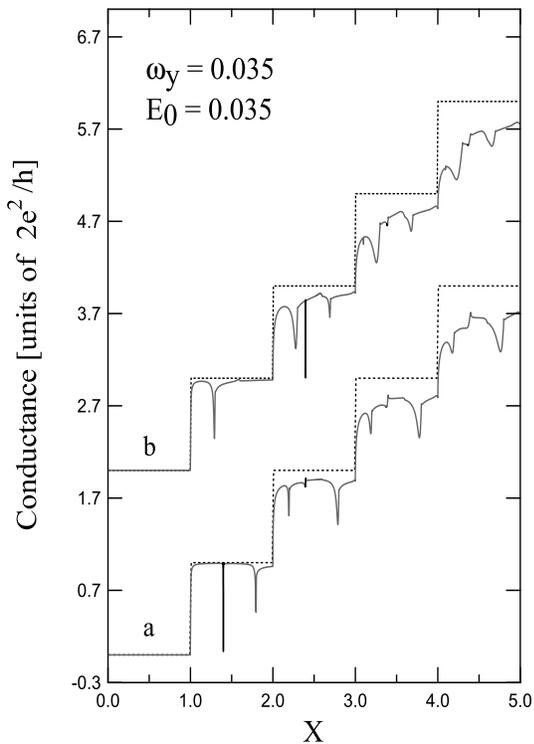}
      \caption{Conductance $G$ as a function of $X$.
$E_{0}=0.0$ for the dotted curves and $E_{0}=0.035$ for the solid
curves.   The frequencies are: (a) $\omega=0.014$ and (b)
$\omega=0.049$.   Curve (b) is vertically offset for clarity.}
\label{fig:1}
\end{figure}

In our numerical examples, the NC is taken to be that in a high
mobility GaAs-Al$_{x}$Ga$_{1-x}$As with a typical electron density
$n \sim 2.5 \times 10^{11} cm^{-2}$,  $m^{*}=0.067 \,m_{e}$, and
$\lambda_{F}=500$\AA.  Correspondingly, our choice of energy unit
$E^{*}=\hbar^{2}k_{F}^{2}/2m^{*}=9 meV$, length unit
$a^{*}=1/k_{F}=79.6$\AA, and frequency unit
$\omega^{*}=E^{*}/\hbar=13.6$THz.  We also take $\omega_{y}=0.035$,
such that the effective channel width is of the order of a few
thousand \AA, and $E_{0}=0.035$.  In Fig.\ 1, we plot $G$ as a
function of the chemical potential.  In Figs.\ 2, 3, and 4, we
present the conductance due to electrons incident from the lowest
subband ($n=0$) and transmitted into various subbands and sidebands.
The analysis for the electrons incident from higher subbands is
similar, though not exactly identical, and is not shown in this
paper.

In Fig.\ 1, $G$ is plotted as a function of $X$, instead of $\mu$,
where $\displaystyle{X=\left[(\mu/\omega_{y})+1\right]/2}$ is the
number of propagating channel.  The dotted curves are the
unperturbed results.  The solid curves in Figs.\ 1(a), and 1(b) have
different electric field frequencies: $\omega=0.014$, and $0.049$,
which correspond to 190 GHz, and 666 GHz, respectively.  The energy
$\hbar\omega$ corresponds to $\Delta X=\omega/(2\omega_{y})$ on the
ordinate. Thus $\Delta X=0.2, \, 0.7$, respectively, in Figs.\ 1(a),
and (b). The two figures show a suppression in $G$, which increases
with $X$.  There are dip structures, dip-and-peak structures in $G$
and the structures are sharper in the smaller $\omega$ region.  The
dip structures around $X=1.8,\ 2.8,\ 3.8,$ and $4.8$ in Fig.\ 1(a)
and around $X=1.3,\ 2.3,\ 3.3,$ and $4.3$ in Fig.\ 1(b), correspond
to the situations when $X+\Delta X=N+1$ are in the neighborhood of
the $N$th subband edge.  At these values of $X$, the electron can
absorb one $\hbar\omega$ and makes an inter-subband transition to
the band bottom of the $N$th  subband which DOS, being proportional
to $\sum_{n=0}^{N}1/k_{n}(\mu+\omega)$, is singular because of the
vanishing $k_{N}(\mu+\omega)$.

It is unexpected, though not entirely surprising, that $G$ exhibits
such a dip-like drop, rather than a spike-like peak, in spite of the
opening up of a new electric-field-assisted transmission channel.
This can be understood in
light of the formation of a QBS.\cite{bag90,bag92}.
The
QBS, formed at energy near a subband edge, traps
temporarily a conduction electron and gives rise to a drop in $G$.
However, due to the interaction of the electron with the electric
field, the trapped electron
can be excited back out of the QBS, hence resulting in a smaller $G$
reduction: $\mid\Delta G\mid<1$, in units of $2e^{2}/h$.  In contrast,
the impurity-induced dips are results of merely elastic scattering and
thus have $G$ reduction $\mid\Delta G\mid=1$.\cite{chu89,bag90}
Our results demonstrate the manifestation of a
new, and electric-field induced, QBS formed from both
inter-subband and inter-sideband transitions.

Other dip structures around $X=2.2,\ 3.2,$ and $4.2$ in Fig.\ 1(a),
and around $X=2.7,\ 3.7,$ and $4.7$ in Fig.\ 1(b) correspond to the
situation $X-\Delta X=N+1$ when the electron can emit one
$\hbar\omega$ and
makes an inter-subband transition to the $N$th subband edge.  Again,
the causes of these structures are the QBS.  Structures
other
than one $\hbar\omega$ processes can also be identified.  The
structures around
$X=1.4,\ 2.4,\ 3.4,$ and $4.4$ in Fig.\ 1(a), and around $X=2.4,\
3.4,$ and
$4.4$ in Fig.\ 1(b) correspond to the situation $X-2\Delta X=N+1$ when
the electron can emit two $\hbar\omega$ and makes transition to the
$N$th subband edge.  We remark, however, that the transitions are
intra-subband
transitions and that the structures, besides the first dip,
do not carry a simple dip-profile.  Rather, they are
dip-and-peak structures which betray the competition between the
trapping of electrons by the QBS and the opening up of a
new electric-field-assisted transmission channel.  The structures in
G due to three $\hbar\omega$ processes are at $X=2.6,\ 3.6,$ and $4.6$
in Fig.\ 1
(a) and at $X=3.1,$ and $4.1$ in Fig.\ 1(b).  These structures are
more
pronounced for larger $\omega$.  Overall, as far as the dependence on
$X$ is concerned, the suppression in $G$ increases as $X$ is
increasing.
In addition, the dip structures and the dip-and-peak structures are
sharper in the smaller $X$ region.

 In the above calculation, for practical purposes, we have to impose a
cutoff to the maximum number of net $\hbar\omega$ involved in the
electron
transmission and the accuracy of our results is checked by the
conservation of current condition, given after Eq.(6).
For our chosen values for $E_{0}$, it is found
that processes involve more than three net $\hbar\omega$
are essentially negligible.  The $m$ net $\hbar\omega$ processes we
refer to are the
processes that the emanating electrons has absorbed or emitted $m$
$\hbar\omega$.
There is, of course, no restriction to the number of
interactions occurring during the transmission of
the electron.  This feature is contained in our results: that we have
incorporated $E_{0}$ and $\omega$ to all orders even for a finite
cutoff to the maximum number of net $\hbar\omega$ processes.
Thus $G$ is found to converge rapidly
with the maximum number of net $\hbar\omega$ processes included.
This is effectively
a renormalization procedure which is deemed necessary due to the
singular DOS near a subband edge.

In conclusion, the QBS features are found in a time-dependent
delta-profile and transversely polarized electric field.  It has been
argued that our results have qualitative implications to the case of a
finite-profile electric field, and hence the photon effects.  Further
study, however, is needed to reach the final conclusion.\\

This work was partially supported by the National Science Council of
the Republic of China through Contract No.\ NSC83-0208-M-009-060.

\end{document}